\documentclass[12pt]{article}
\usepackage{graphicx}
\usepackage[margin=1.25in]{geometry}
\usepackage[usenames,dvipsnames]{color}
\usepackage{url}
\usepackage[colorlinks = true,
            linkcolor = blue,
            urlcolor  = blue,
            citecolor = blue,
            anchorcolor = blue]{hyperref}
\usepackage{lineno}            

\newcommand\pubnumber
{DESY 17-237\\
 KEK Preprint 2017-57\\
 SLAC-PUB-17197\\}
\newcommand\pubdate{\today}

\def\KEK{High Energy Accelerator Research Organization (KEK), Tsukuba,
  Ibaraki, JAPAN  }
\def\Tokyo{ICEPP, University of Tokyo, Hongo, Bunkyo-ku, Tokyo,
  113-0033, JAPAN}
\def\SNU{Dept. of Physics and Astronomy, Seoul National
  Univ.,  Seoul 08826, KOREA}
\def\DESY{DESY, Notkestrasse 85, 22607 Hamburg, GERMANY}
\def\DESYZ{DESY, Platanenallee 6, 15738 Zeuthen, GERMANY}
\def\Berlin{Institut f\"ur Physik, Humboldt-Universit\"at zu Berlin, 12489 Berlin, GERMANY}
\def\SLAC{SLAC,
    Stanford University, Menlo Park, CA 94025, USA}
\def\Tsinghua{Center for High Energy Physics, Tsinghua University,
  Beijing, CHINA}
\def\Osaka{Department of Physics, Osaka University, Machikaneyama, Toyonaka, Osaka 560-0043, JAPAN}
\def\IPMU{Kavli Institute for the Physics and Mathematics of the Universe,
University of Tokyo, Kashiwa 277-8583, JAPAN}
\def\Cornell{Laboratory for Elementary Particle Physics, Cornell
  University, Ithaca, NY 14853, USA}
\def\Orsay{LAL, Centre Scientifique d'Orsay, Universit\'e Paris-Sud, F-91898 Orsay CEDEX,
FRANCE}
\def\Munich{Max-Planck-Institut f\"ur Physik, F\"ohringer Ring 6,
  80805 Munich, GERMANY}
\def\Michigan{Michigan Center for Theoretical Physics, University of Michigan, Ann Arbor,
MI 48109, USA}
\def\UTA{Department of Physics, University of Texas, Arlington, TX
  76019, USA}
\def\Oregon{Center for High Energy Physics, University of Oregon, Eugene, Oregon
97403-1274, USA}
\def\Berkeley{ Department of Physics, University of California, Berkeley, CA 94720, USA}
\def\LBNL{Theoretical Physics Group, Lawrence Berkeley National Laboratory, Berkeley,
CA 94720, USA}
\def\UHamburg{Department of Physics, Hamburg University, Luruper Chausee 149, 22761 Hamburg, GERMANY}
\def\Kansas{Department of Physics and Astronomy, University of Kansas, Lawrence, KS 66045, USA}

\def\Title#1{\begin{center} {\Large #1 } \end{center}}
\def\Author#1{\begin{center}{ \sc #1} \end{center}}

\newcommand\pubblock{\rightline{\begin{tabular}{l} \pubnumber\\
         \pubdate \end{tabular}}}
\newenvironment{Abstract}{\begin{quotation} \begin{center}
                       ABSTRACT
     \end{center}\bigskip  }{\end{quotation}}

\def\Acknowledgements{\bigskip  \bigskip \begin{center} \begin{large}
             \bf ACKNOWLEDGEMENTS \end{large}\end{center}}

\def\beq{\begin{equation}}
\def\eeq#1{\label{#1}\end{equation}}
\def\eeqn{\end{equation}}

\newenvironment{Eqnarray}%
   {\arraycolsep 0.14em\begin{eqnarray}}{\end{eqnarray}}
\def\beqa{\begin{Eqnarray}}
\def\eeqa#1{\label{#1}\end{Eqnarray}}
\def\eeqan{\end{Eqnarray}}

\let\bar=\overbar

\def\lsim{\mathrel{\raise.3ex\hbox{$<$\kern-.75em\lower1ex\hbox{$\sim$}}}}
\def\gsim{\mathrel{\raise.3ex\hbox{$>$\kern-.75em\lower1ex\hbox{$\sim$}}}}

\def\L{{\cal L}}

\def\L{{\cal L}}
\def\P{{\cal P}}

\def\del{\partial}
\def\Dslash{\not{\hbox{\kern-4pt $D$}}}
\def\dslash{\not{\hbox{\kern-2pt $\del$}}}

\def\MET{\not{\hbox{\kern-4pt $E$}}_T}

\def\Dlr{\mathrel{\raise1.5ex\hbox{$\leftrightarrow$\kern-1em\lower1.5ex\hbox{$D$}}}}

\def\eff{{\mbox{\scriptsize eff}}}

\def\ee{e^+e^-}

\def\msb{{\bar{\scriptsize M \kern -1pt S}}}

\def\drb{{\bar{\scriptsize D \kern -1pt R}}}

\def\Leff{\ensuremath{\L_{\eff}}}
\def\Peff{\ensuremath{\P_{\eff}}}
\def\Pe{\ensuremath{\P_{e^-}}}
\def\Pp{\ensuremath{\P_{e^+}}}
\def\Ppe{\ensuremath{\P_{e^{\pm}}}}
\def\Pep{\ensuremath{\P_{e^{\mp}}}}

\makeatletter
\def\section{\@startsection{section}{0}{\z@}{5.5ex plus .5ex minus
 1.5ex}{2.3ex plus .2ex}{\large\bf}}
\def\subsection{\@startsection{subsection}{1}{\z@}{3.5ex plus .5ex minus
 1.5ex}{1.3ex plus .2ex}{\normalsize\bf}}
\def\subsubsection{\@startsection{subsubsection}{2}{\z@}{-3.5ex plus
-1ex minus  -.2ex}{2.3ex plus .2ex}{\normalsize\sl}}

\renewcommand{\@makecaption}[2]{%
   \vskip 10pt
   \setbox\@tempboxa\hbox{\small #1: #2}
   \ifdim \wd\@tempboxa >\hsize     
       \small #1: #2\par          
     \else                        
       \hbox to\hsize{\hfil\box\@tempboxa\hfil}
   \fi}

 \def\citenum#1{{\def\@cite##1##2{##1}\cite{#1}}}
 
\newcount\@tempcntc
\def\@citex[#1]#2{\if@filesw\immediate\write\@auxout{\string\citation{#2}}\fi
  \@tempcnta\z@\@tempcntb\m@ne\def\@citea{}\@cite{\@for\@citeb:=#2\do
    {\@ifundefined
       {b@\@citeb}{\@citeo\@tempcntb\m@ne\@citea\def\@citea{,}{\bf ?}\@warning
       {Citation `\@citeb' on page \thepage \space undefined}}%
    {\setbox\z@\hbox{\global\@tempcntc0\csname b@\@citeb\endcsname\relax}%
     \ifnum\@tempcntc=\z@ \@citeo\@tempcntb\m@ne
       \@citea\def\@citea{,}\hbox{\csname b@\@citeb\endcsname}%
     \else
      \advance\@tempcntb\@ne
      \ifnum\@tempcntb=\@tempcntc
      \else\advance\@tempcntb\m@ne\@citeo
      \@tempcnta\@tempcntc\@tempcntb\@tempcntc\fi\fi}}\@citeo}{#1}}
\def\@citeo{\ifnum\@tempcnta>\@tempcntb\else\@citea\def\@citea{,}%
  \ifnum\@tempcnta=\@tempcntb\the\@tempcnta\else
  {\advance\@tempcnta\@ne\ifnum\@tempcnta=\@tempcntb \else\def\@citea{--}\fi
    \advance\@tempcnta\m@ne\the\@tempcnta\@citea\the\@tempcntb}\fi\fi}
\makeatother

\def\met{\not{\hbox{\kern-4pt $E$}}_T}

\usepackage{cite, caption, subcaption}

\begin{document}
\begin{titlepage}
\pubblock

\vfill
\Title{The role of positron polarization for the inital $250$\,GeV  stage of the International Linear Collider}
\vfill 
 \Author{LCC Physics Working Group}

\bigskip

\bigskip

\Author{Keisuke~Fujii$^1$, Christophe~Grojean$^{2,3}$, 
  Michael~E.~Peskin$^4$  (Conveners); Tim~Barklow$^4$, Yuanning~Gao$^5$,
  Shinya~Kanemura$^6$, Hyungdo~Kim$^7$, Jenny~List$^2$, 
  Mihoko~Nojiri$^{1,8}$, Maxim~Perelstein$^{9}$, Roman~P\"oschl$^{10}$,
  J\"urgen~Reuter$^{2}$, Frank~Simon$^{11}$, Tomohiko~Tanabe$^{12}$,
  James~D.~Wells$^{13}$, Jaehoon~Yu$^{14}$;  Mikael~Berggren$^{2}$, 
  Moritz~Habermehl$^{2,15}$,
  Robert~Karl$^{2,15}$, Gudrid~Moortgat-Pick$^{15}$, 
  Sabine~Riemann$^{16}$, 
  Junping~Tian$^{12}$, Graham~W.~Wilson$^{17}$;
  James~Brau$^{18}$, Hitoshi~Murayama$^{8,19,20}$ (ex officio)}

\vfill
\begin{Abstract}
The International Linear Collider is now proposed with a staged
machine design, with the first stage at $\sqrt{s}=$~250\,GeV and an integrated luminosity 
goal of 2~ab$^{-1}$. One of the questions for the machine design 
is the importance of positron polarization.  In this report, we review the impact of positron polarization on the physics goals of the $250$\,GeV stage of the ILC and demonstrate that positron polarization has distinct advantages.

\end{Abstract}
\vfill

\end{titlepage}

\noindent $^1$  \KEK\\
$^2$  \DESY\\
$^3$  \Berlin\\
$^4$  \SLAC \\ 
$^5$   \Tsinghua\\
$^6$  \Osaka\\ 
$^7$  \SNU \\ 
$^8$  \IPMU \\
$^9$  \Cornell \\
$^{10}$  \Orsay\\
$^{11}$  \Munich \\ 
$^{12}$   \Tokyo \\
$^{13}$  \Michigan\\ 
$^{14}$  \UTA \\ 
$^{15}$  \UHamburg \\
$^{16}$   \DESYZ \\
$^{17}$  \Kansas \\
$^{18}$   \Oregon\\  
$^{19}$  \Berkeley\\
$^{20}$   \LBNL\\


\hbox to\hsize{\null}

\newpage

\section{Introduction}

Recently, the plan for the International Linear Collider (ILC) has been revised to a staged machine design with the first stage at $\sqrt{s}=250$\,GeV~\cite{Evans:2017rvt}. The physics for such a staged machine has been sumarized in~\cite{Fujii:2017vwa}, based on electron and positron beam polarization, with 80\% polarization of the electron beam and 30\% polarization of the positron beam, as foreseen in the ILC  Technical Design Report~\cite{Adolphsen:2013kya}. Electron polarization is essential for all of the physics goals of the ILC.  It plays an important role in the measurements proposed for every physics topic that will be studied at this machine~\cite{ILCTDR_VOL2}, and we will not comment further on the role of electron polarization in this document.  

The baseline design of the $250$\,GeV stage of the ILC described in~\cite{Evans:2017rvt} includes a polarized positron source. However, because there exist alternative concepts with complementary strengths and weaknesses, the Linear Collider Collaboration (LCC) requested a survey of the importance of positron polarization to meet the physics goals of the $250$\,GeV stage. 
This report is intended to address that request.

The role of positron polarization at future $e^+e^-$ colliders has been reviewed 
in great detail in the past~\cite{MoortgatPick:2005cw}, and  updated for the case of $30$\% positron polarization~\cite{POWER_SB2009}. These reports identified three main benefits of positron polarization.  In this report, we will trace the influence of these through the physics topics of the $250$\,GeV stage of the ILC.

There are three main effects of positron beam polarization which will be discussed in the context of specific physics examples in this note:
\begin{enumerate}
\item 
Positron polarization allows us to obtain subsamples of the data with higher rates for interesting physics processes and lower rates for backgrounds. Since  sensitivities do not combine as a linear sum,
the combination of results from e.g.\ two data sets with small and large signal-to-background ratio, respectively, is more sensitive than a single data set with the same total number of signal and background events. 
\item Positron polarization offers four distinct data sets instead of the two available if only the electron beam can be polarized. Most important reactions can be studied with the opposite-sign polarization modes only, but there are measurements in which the two like-sign polarization states give additional or even unique information. The flexibility in choosing between these configurations (and possibly even five more when considering parts of the data to be taken with zero longitudinal polarisation) is a unique asset of the ILC.  
\item The likely most important effect is the control of systematic uncertainties: The precisions aimed for at the ILC can only be reached if all relevant systematic uncertainties are controlled to the same level as the statistical uncertainties or better. This requires sufficient experimental redundancy in order to determine all relevant nuisance parameters in-situ. While it is difficult to estimate reliably systematic uncertainties in the absence of real detectors and real data, we present one example based on ILC Monte-Carlo studies below. In addition we discuss a real-life example from the SLC which is very instructive also for the ILC.
\end{enumerate}

As we survey the various physics topics, we will conclude that the first of these advantages can in many cases be compensated by an increase of the running time by $2-3$ years. The second advantage is a qualitative one,
and the loss in degrees of freedom from $4$ to $2$ independent data-sets\footnote{or, when including the case of zero longitudinal polarisation by either rotating to transverse polarisation or by active depolarisation from $9$ to $3$ data-sets} will be paid in terms of less model-independence of the measurements and their interpretation.  

The third advantage that we have cited here is less easily quantified.  However, the advantage of positron polarization for systematic control in precision experiments is a very important one for the ILC program. In particular in the event that the ILC discovers an anomaly with respect to the Standard Model in a precision observable, the additional measurements made possible by positron polarization will be important to give confidence in the presence of this effect with respect to possible systematic uncertainties.

On this basis, we find that positron polarization has an important role to play in the ILC program.  


This note is structured as follows: we will begin with a recap of polarization formalism in Section~\ref{sec:polbasics} and discuss polarimetry, with special emphasis on the case \Pp=0 in Section~\ref{sec:polarimetry}. In Section~\ref{sec:sm}, we will illustrate the importance of positron polarization for the control of systematic uncertainties by the example of 2- and 4-fermion processes including $W$ boson pair production, and explain the role of positron polarization in the search for new sources of CP violation. Section~\ref{sec:higgs} addresses the influence of positron polarization on precision Higgs boson measurements. Section~\ref{sec:bsm}  discusses positron polarization in the context of searches for new phenomena. We finally give our conclusions in Section~\ref{sec:conclusions}.

\section{Polarization at $\ee$ Colliders}
\label{sec:polbasics}

In order to discuss the effects listed above in more detail, we recall a few general considerations. As weak interactions are chiral, i.e.\ $W^{\pm}$ and $Z$ bosons couple differently to left-handed and right-handed fermions, and only left-handed fermions take part in charged weak interactions, polarization effects play an important role in strategies to extract information from $\ee$ reactions. Since all the strong motivations to search for physics beyond the Standard Model (SM) of particles physics relate to its electroweak sector, there is every reason to expect that also new phenomena will depend on the chirality of the involved particles.

Any real particle beam will contain a mixture of $N_L$ left- and $N_R$ right-handed particles, given by the longitudinal beam polarization
\begin{equation}
\P = \frac{N_R - N_L}{N_R + N_L}
\end{equation}
Depending on the orientation of the full polarization vector, there can also be transverse polarization, which will play a role later when we discuss CP violating effects. The spin rotator systems of the ILC~\cite{Malysheva:2016jdr}, which are needed to turn the polarization into the vertical before the damping rings and back into the longitudinal direction afterwards, have been designed carefully to allow 
any orientation of the polarization vectors at the $\ee$ interaction point. In the following, however, ``polarization'' refers by default to longitudinal polarization, unless transvese polarization is explicitly mentioned.

For electron beam polarizations \Pe\ and positron beam polarization \Pp , the cross section of any reaction is computed from the four possible pure chiral cross sections (with $\sigma_{\rm LR}$ for left-handed electron and right-handed positron etc.) as
\begin{eqnarray}
\sigma(\Pe,\Pp) &=& \frac{1}{4}\bigl\{
  (1+\Pe)(1+\Pp)\sigma_{\rm RR} 
+ (1-\Pe)(1-\Pp)\sigma_{\rm LL} \nonumber \\
&& + (1+\Pe)(1-\Pp)\sigma_{\rm RL} 
+ (1-\Pe)(1+\Pp)\sigma_{\rm LR} \bigr\},
\end{eqnarray}
The unpolarized cross section $\sigma_0$ is given by
\begin{equation}
\sigma_0 = \frac{1}{4}\bigl\{\sigma_{\rm RR} + \sigma_{\rm LL} + \sigma_{\rm RL} + \sigma_{\rm LR} \bigr\}.
\end{equation}
Further important quantities are the left-right asymmetry $A_{\rm LR}$, the effective luminosity \Leff, and the effective polarization \Peff:
\begin{eqnarray}
A_{\rm LR} &=& \frac{(\sigma_{\rm LR} - \sigma_{\rm RL})}{(\sigma_{\rm LR} + \sigma_{\rm RL})} \\
\Leff &=& \frac{1}{2} (1-\Pe \Pp ) \L \\
\Peff &=& \frac{\Pe -\Pp}{1 - \Pe \Pp}.
\end{eqnarray}
The quantity \Peff\ can be substantially closer to $\pm 1$ than achievable single beam polarizations. For example, for $\Pp = +0.3$ and $\Pe = -0.8$, $\Leff = 0.62 \L$ and $\Peff = -0.89$, while in case of $\Pp = 0$, $\Leff = 0.5 \L$ and $\Peff = \Pe$.
The drop in $\Leff$ means that descoping from \Pp=30\% to \Pp=0 is for most processes equivalent to a 24\% loss in luminosity, and at the same time the reduction of $\Peff$ translates into a 10\% reduction of analysing power for left-right asymmetries, which are important observables for electroweak and Higgs physics.

For the ILC physics programme, several distinct types of process are of particular importance:
\begin{itemize}
\item {\bf $s$-channel $Z/\gamma$ exchange:} For the $s$-channel exchange of a vector boson, the spin of the incoming particles have to add up to a total spin-1 configuration, therefore only $\sigma_{LR}$ and $\sigma_{RL}$ are non-zero. In this case, the polarized cross section simplifies to:
\begin{equation}
\sigma(\Pe,\Pp) = 2 \sigma_0 (\Leff / \L) [1- \Peff A_{\rm LR}]
\end{equation}
Important examples are Higgs production via Higgs-strahlung and fermion-anti-fermion production, e.g.\ $b\bar{b}$ production.
\item{\bf $t$-channel $W$ or $\nu_e$ exchange:} Since only left-handed fermions and right-handed anti-fermions take part in the charged weak interaction, only $\sigma_{LR}$ is non-zero in this case, or in other words $A_{\rm LR} = 1$, thus 
\begin{equation}
\sigma(\Pe,\Pp) = 2 \sigma_0 (\Leff / \L) [1 - \Peff]
\end{equation}
Important examples are Higgs production via $WW$ fusion and $W$ pair production, but also neutrino pair production as background for missing energy signatures.
In this case, the impact of \Leff\ and \Peff\ is even more striking: for $\Pp = +0.3$ and $\Pe = -0.8$, the cross section increases by 30\% w.r.t.\ $\Pp = 0$ and $\Pe = -0.8$. In the opposite sign configuration, which in case of e.g.\ $WW$ measurements serves as in-situ background determination,  $\Pp = -0.3$ and $\Pe = +0.8$ gives a 30\% reduction of the remaining signal pollution w.r.t.\ $\Pp = 0$ and $\Pe = +0.8$.
\item{\bf single $W$ production:} In this case, only one of the beam particles emits a $W$ boson, which scatters on a photon (or a $Z$) from the other beam. Since the photon and the $Z$ boson couple to left- and right-handed fermions, $\sigma_{LR}$ and $\sigma_{LL}$ are allowed for $W^-$ production, while $\sigma_{LR}$ and $\sigma_{RR}$ are responsible for $W^+$ production. So in these cases, the polarized cross sections are given by:
\begin{equation}
\sigma^{\pm}(\Pe,\Pp) = \sigma_0 (1 \pm \Ppe) [1 - \Pep A_{\rm RR/LL}]
\end{equation}
with the upper signs for $W^+$ production and the lower ones for $W^-$ production. The relevant asymmetries are $A_{\rm RR} = (\sigma_{\rm LR} - \sigma_{\rm RR})/(\sigma_{\rm LR} + \sigma_{\rm RR})$ and $A_{\rm LL} = (\sigma_{\rm LR} - \sigma_{\rm LL})/(\sigma_{\rm LR} + \sigma_{\rm LL})$, respectively.
\item{\bf new physics:}
Depending on the kind of new physics, in principle all four chiral cross sections can be relevant.
The measurement of the full set of chiral cross sections delivers important information on the interaction of the new particles with SM particles. We will discuss WIMP dark matter as a specific example below. 
\end{itemize}

In order to evaluate the role of positron polarization quantitatively, we need to refer to a specific 
operation scenario. Throughout this document, we will assume a total integrated luminosity of $2$\,ab$^{-1}$ to be collected at a center-of-mass energy of $250$\,GeV. For the case of $|\Pp| = 30\%$, we assume that $45\%$ of the data will be collected with $\Pe = -80\%$ and $\Pp = +30\%$, another $45\%$ with reversed signs, and $5\%$ on each of the like-sign configurations, in accordance with~\cite{Fujii:2017vwa}. In the absence
of positron polarization, we assume half of the data to be taken with $\Pe = -80\%$, and the other half with $\Pe = +80\%$. These specific combinations will be referred to as ``\Pp = 30\%'' and ``\Pp = 0'' scenarios in the remainder of the document.

\section{Determination of the Beam Polarization}
\label{sec:polarimetry}
The polarimeters~\cite{Boogert:2009ir} in the beam delivery system of the ILC will provide fast online measurements of the beam polarization  $\sim 1.7$\,km before and $\sim 150$\,m behind the $e^+e^-$ interaction point with precisions of $0.25\%$. However the luminosity-weighted average polarization during the $e^+e^-$ collisions, which can differ from the polarimeter measurements due to spin transport and depolarization effects~\cite{Beckmann:2014mka}, will ultimately be obtained by a combination of the polarimeter measurements with $e^+e^-$ cross section measurements. For the case of \Pp = 30\% it has recently been shown that the goal of 0.1\% precision on the luminosity-weighted average polarization can be reached~\cite{Karl:2017xra} for each of the datasets foreseen in the H-20 running scenario~\cite{Barklow:2015tja}, even when treating the absolute values of the beam polarizations for positive and negative signs as independent parameters, thus allowing for an imperfect helicity reversal.

Before we discuss the effect of the positron polarization on Higgs and SM measurements as well as on BSM searches in the next sections, it is important to address the following question: Can we, in the case of an unpolarized positron source, rely under all circumstances on $\Pp \equiv 0$ without in-situ control of this assumption? Or should \Pp\ rather be included as nuisance parameter, possibly constrained within the polarimeter uncertainty?

While there is no obvious origin of polarization (other than the polarized sources) in the design of the ILC, we need to consider what
would happen if we indeed observe a discrepancy from the Standard Model prediction, e.g.\ in some cross-section asymmetry. Before being able to claim an observation of physics beyond the SM, it would be our duty to exclude any other explanation for the observed discrepancy, however unlikely it might seem. Depending on how well this will then be possible {\em a posteriori}, any remaining doubt would shadow the ILC's discovery potential.

Actually there is an example of exactly such a situation in the measurement of the effective weak mixing angle via $A^e_{\rm LR}$ by the SLD experiment and via $A^b_{\rm FB}$ at LEP, which differ by more than $3\,\sigma$~\cite{ALEPH:2005ab}. While the SLD measurement of course relied on the SLC electron beam being polarized, the SLC positron beam was nominally unpolarized and not equipped with a polarimeter. In view of the above mentioned discrepancy, the SLD collaboration in the end undertook a considerable effort in order to measure the positron polarization {\em a posteriori} to a precision of $\pm 0.0007$ in a dedicated experiment at SLAC's End Station A~\cite{SLDnote268} --- despite the fact that nobody had a mechanism for producing non-zero positron polarization from the SLC positron target!

We conclude from this lesson than even for the \Pp = 0 scenario, the 
positron polarization should be treated as a nuisance parameter in global fits, and that the polarimeters for the positron beam are essential even if the nominal polarization is zero.

\section{Standard Model Precision Measurements}
\label{sec:sm}

Precision measurements of all kinds of electroweak observables are at the heart of the ILC's physics program, and will be the basis for the precise and model-independent characterization of the Higgs bosons. Observables like total and differential cross sections, left-right and forward-backward asymmetries etc.\ enable us to probe energy scales far beyond the center-of-mass energy of the collider. See, e.g.,~\cite{Fujii:2017vwa} and references therein.

Higher precision of the measurements will thereby allow to probe higher energy scales. In order to reduce the impact of systematic uncertainties
to a minimum, it will be necessary to constrain the actual observables of interest simultaneously with many so-called nuisance parameters, which model possible systematic effects. Classic examples comprise the
luminosity, the beam polarizations, selection efficiencies as well as theoretical or parametric uncertainties.

\subsection{Cross section and asymmetry measurements}
\label{sec:sm:xs}
Only recently a study to demonstrate the simultaneous extraction of total cross sections, left-right asymmetries and beam polarizations from differential distributions of all kinds of electroweak processes at the ILC was started. The results will be reported in this document for the first time, based on an extension of the framework described in~\cite{Karl:2017xra} for the beam polarization extraction\footnote{
A further extension of this framework to include also anomalous triple gauge couplings is in progress.}. This study is currently by far the most comprehensive attempt at such a global interpretation, and directly targets ILC operation at 250\,GeV.


The study makes use of the errors projected for ILC on the following observables:  for $W$ pair production, single W production, the total cross sections (for each setting of the beam polarizations) and the (binned) differential cross sections with respect to the $W$ production and decay angles;  for single $W$ production, the total cross sections and differential cross sections with respect to the $W$ decay angles; for 2-fermion processes, the total cross sections and differential cross sections with respect to the fermion production angle.  Measurements of the beam polarizations by the polarimeters to an accuracy of 0.0025 are also included.  The analysis also includes the possibility of an undetected bias between the polarization measured at the IP. A fit is performed and the following parameters are extracted:   the total unpolarized cross section for each process, the left-right asymmetry for each process, and the beam polarizations separately for $e^-$ and $e^+$, a total of 16 parameters.  In the case \Pp=0, only one parameter is considered for the positron polarization, giving 15 parameters in total.

While a full description of the procedure and it's results would go far beyond the scope of this document, we summarize here the findings which are of highest relevance to the impact of positron polarization. A full documentation will be available in early 2018 in the PhD thesis of R.~Karl~\cite{Karl_thesis}. All numbers in the following are lower limits since detector inefficiencies and instrumental backgrounds have not yet been included. We therefore stress most the {\em relative} changes between different configurations, which are not expected to be significantly affected by these simplifications.

\begin{itemize}
\item In all the configurations studied so far, the electron polarization was always well determined to sub-per mille precision.
\item Without a constraint on the positron polarization from the polarimeters, the positron polarization can only be constrained to
0.5\% in the \Pp=0 case. This is a factor 5 worse than usually assumed as systematic uncertainty, e.g., in the Higgs boson studies described in Sec.~\ref{sec:higgs}. At the same time, the resulting uncertainties on the total cross sections and on the left-right asymmetries grow by typically one order of magnitude compared to the case of \Pp=30\%. 
\item When adding the polarimeter constraint for the positron polarization, assuming no bias, the effects are partially mitigated:
The polarimeter uncertainty propagates nearly one-to-one onto the positron polarization obtained from the fit, so with the polarimeter measuring $\Pp = 0 \pm 0.0025$, the final uncertainty is still 0.0024.
The same applies for the total cross sections for $W$ pair and single-$W$ processes and the left-right asymmetries for the 2-fermion processes, which still are a factor 2-3 worse than in the \Pp=30\% case. 
\item If now a bias of 2.5 per mille is assumed between the polarimeter and the IP, thus  $\Pp = 0.0025 \pm 0.0025$, we observe the following effects: In the case of \Pp = 30\%, the fitted cross sections and asymmetries receive a bias of typically $-0.5 \sigma$, thus covered by the uncertainty. A larger discrepancy could be revealed by comparison with the results from a fit without the polarimeter constraint, which in case of \Pp = 30\% is, for sufficiently large data-sets, of similar precision. In the case of \Pp = 0, however, all total cross sections and three out of the six considered asymmetries 
receive biases between $-1$ and $-1.5\,\sigma$. Since the uncertainties without the polarimeter constraint are an order of magnitude larger, as described in the previous bullet, the opportunity of an independent consistency check does not exist in the case of \Pp = 0.
\end{itemize}

Although this study is still work in progress, it illustrates that positron polarization plays an important role in beating down systematic uncertainties for all kinds of cross section and asymmetry measurements --- including Higgs observables.

\subsection{$CP$ violation}
\label{sec:sm:cp}
$CP$ violation is one of the key ingredients needed to explain the baryon-antibaryon asymmetry of the universe, which is fundamental to our existence. It is  well established that the $CP$ violation observed in the quark sector of the SM is too small to explain this asymmetry. Thus, we hope to discover additional sources of $CP$ violation in the Higgs sector or in the neutrino sector. This is among the most important motivations for upcoming and future experiments.

Observables such as total and differential cross sections are $CP$-invariant, and so the observation of $CP$ violation requires additional observables beyond those discussed in the previous section.
In physics channels with sufficiently complex final states,  $CP$-sensitive triple products can be constructed, or $\tau$ polarization can be exploited. In other cases, however, initial state
polarization, and here especially transverse polarization is the only possibility. While it is not part of the 
default ILC operating scenario, the spin rotators could provide transverse polarization. The ILC polarimeters can measure this transverse polarisation when equipped with an appropriate detector for the Compton-scattered electrons~\cite{Mordechai:2013zwm}. For measurements sensitive to longitudinal polarization only, data-sets  with transverse polarizations are equivalent to data setz with unpolarized beams, just as it was the case at LEP~\cite{Arnaudon:1992rn}, or at HERA~\cite{Kramer:1994cv} before the installation of spin rotators. 

The importance of positron polarization in searches for $CP$ violation in di-boson production depends on the number of parameters used to describe cross sections.   For the process $\ee \to W^+W^-$, an Effective Field Theory description with dimension-6 operators only gives 5 parameters to describe the new physics effects. Of these, 3 are $CP$-conserving and 2 are $CP$-violating.  All of these parameters can be extracted using only longitudinal polarization.  Positron polarization is not required, though it gives a quantitative advantage in the statistical precision~\cite{Fujii:2017vwa,Marchesini:2011aka,Rosca:2016hcq}.  Similarly, in $\ee\to \gamma Z$, the EFT description gives only one $CP$-violating parameter, which can be measured even without any beam polarization.

However, if we would like to test the most general parameter set for these reactions, the situation is different.  Assuming only Lorentz invariance, the triple gauge boson vertices in  $\ee\to W^+W^-$ allows 14 complex (or 28 real) parameters~\cite{Diehl:2003qz}. Several of these parameters require positron polarization for their determination.  In particular, the $CP$ violating parameter $h^+ = \mbox{Im}(g^R_1 + \kappa^R_1)/\sqrt{2}$ appears in the cross section formula in a term that contains the product of \Pp\ and \Pe\ and so explicitly requires both beams to be (transversely) polarized. Similar arguments apply to the determination of the most general parameters for the triple gauge boson vertices contributing to $\ee\to \gamma Z$~\cite{MoortgatPick:2005cw,POWER_SB2009}. Measurements based on these most general parameter sets are only possible with positron polarization.

%

\section{Precision Characterization of the 125-GeV Higgs Boson}
\label{sec:higgs}
The most obvious effect of positron polarization on Higgs physics is an increase of the number of produced Higgs bosons
by about $20\%$: Assuming SM cross sections, about 420 000 Higgs bosons will be produced in the \Pp=0 scenario,
while this number increases to about 500 000 in the \Pp=30\% case. These numbers are based on the luminosity sharing between helicity configurations as defined at the end of Section~\ref{sec:polbasics}.  Thus without positron polarization, the running time (and thus the running costs) would need to be increased by 19\% in order to reach the same number of produced Higgs bosons.  

The report~\cite{Barklow:2017suo} presented projections of uncertainties in Higgs boson couplings using an analysis based on Effective Field Theory.  That analysis was based on the \Pp=30\% scenario. The results of the same analysis redone for the \Pp=0 case and also with both polarizations zero are shown in Table~\ref{tab:EFThiggs}. In this 
interpretation, the loss of positron polarization has only a relatively small effect, degrading the Higgs coupling ratios $g(h\tau\tau)/g(hWW)$ and $g(hbb)/g(hWW)$ by $6\%$, with smaller effects on the absolute coupling determinations. This is mainly a statistical effect consistent with the drop in $\Leff$ explained in Section~\ref{sec:polbasics}, since coupling precisions scale inversely with the square root of the luminosity.
\begin{table}[h]
\begin{center}
\begin{tabular} {lcccccc}
 & no pol.   & 80\%/0\%    & 80\%/30\% \\
\hline
$g(hbb)$                           &    1.33        &  1.13        & 1.09  \\
$g(hcc)$                            &    2.09        &  1.97        & 1.88  \\
$g(hgg)$                           &    1.90        &  1.77        & 1.68  \\
$g(hWW)$                        &    0.978      &  0.683      & 0.672 \\
$g(h\tau\tau)$                   &    1.45        &  1.27        & 1.22   \\
$g(hZZ)$                          &    0.971       &  0.693     & 0.682  \\
$g(h\gamma\gamma)$     &    1.38        &  1.23        & 1.22   \\
$g(h\mu\mu)$                   &    5.67        &  5.64        & 5.59   \\
$g(h\gamma Z)$               &    14.0        &  6.71        & 6.63   \\
\hline
$g(hbb)/g(hWW)$             &    0.911      &  0.909       & 0.861 \\
$g(h\tau\tau)/g(hWW)$     &    1.08        &  1.08         & 1.02   \\
$g(hWW)/g(hZZ)$            &     0.070      &  0.067       & 0.067 \\
\hline
$\Gamma_h$                   &    2.93         &  2.60         & 2.49   \\
\hline
$BR(h\to inv)$                  &    0.365       &  0.327       & 0.315  \\
$BR(h\to other)$              &    1.68          &  1.67         & 1.58   \\
\end{tabular}
\caption{Projected relative errors for Higgs boson couplings and other Higgs observables at 250 GeV, in \%,
comparing three cases of beam polarization: $2$\,ab$^{-1}$ with \Pe = \Pp = 0\%, as well as the \Pp=0 and \Pp=30\% scenarios defined in the Introduction.}
\label{tab:EFThiggs}
\end{center}
\end{table}

However it should be noted that these results were obtained by scaling only the statistical uncertainties and assuming that all the systematic uncertainties stay the same, independently of the polarization. In particular
it was assumed that the uncertainties on the measured cross sections ($\times BR$) due to finite knowledge of the luminosity and the polarization are $0.1\%$ each, whereas for $H\to b\bar{b}$-channels an additional uncertainty of $0.1\%$ on the $b$-tagging efficiency has been considered. No uncertainties on e.g.\ the residual background contributions etc.\ were taken into account. While it is clearly better than not including any systematic uncertainties at all, this scheme does not reflect the increase in systematic uncertainties expected for the \Pp=0 case:
\begin{itemize}
\item {\bf luminosity uncertainty}: As we discussed for the example of SM cross section and asymmetry measurements in Section~\ref{sec:sm:xs}, the precision with which a global scaling uncertainty can be pinned down by treating it as a nuisance parameter in a global fit of many observables can depend quite significantly on the presence or absence of positron polarization.
\item {\bf polarization uncertainty}: Likewise we showed in Section~\ref{sec:sm:xs} that the precision to which the positron polarization can be determined in-situ depends significantly on its absolute value. For \Pp=0, it will be given de facto by the positron polarimeter measurement alone, and any bias in the polarimeter measurement
will propagate into $\ee$ cross section measurements.
\item {\bf $b$-tagging uncertainty}: It still needs to be investigated in the future in how far the $b$-tagging efficiency could be constrained better
from data when using all four datasets which are available in the \Pp = 30\% scenario, similar to the EW cross section example given below.
\item {\bf background uncertainty}: While such types of uncertainty are not included at all in the current Higgs coupling fit, it has been shown in other contexts, e.g.\ the $WW$-threshold scan~\cite{Wilson:2016hne}, that the datasets which have beam helicity configurations unfavourable for the signal can give a very important handle on the residual background. For many Higgs channels, $W$-pair production is an important background. In principle $WW$ and $ZH$ production exhibit already very different polarization dependence with electron polarization only.
It should be subject of future studies whether this gives enough redundancy to control systematic uncertainties, since actually the left-right asymmetry of the $ZH$ cross section plays an important role as input observable in EFT-based fits~\cite{Barklow:2017suo,Durieux:2017rsg}. Therefore, positron polarization is needed here as an extra handle in order to simultaneously constrain the left-right asymmetry of the signal and background contributions with different polarization. 
\end{itemize}

Finally, though the EFT-based Higgs coupling fit has strong theoretical justification, it would also be desirable to test the validity of this framework from observations.  Some parameters in this fit are constrained by precision electroweak measurements and constraints on anomalous triple gauge couplings. In~\cite{Barklow:2017suo}, the triple gauge couplings are constrained to depend on only 3 parameters (corresponding to shifts in $g_Z$, $\kappa_{\gamma}$ and $\lambda_{\gamma}$), as predicted by the EFT formalism in leading order. As we have discussed in Section~\ref{sec:sm:cp}, it is possible at the ILC to constrain the most general set of triple gauge coupling deviations allowed by Lorentz invariance, but only if both polarized electron and positron beams are available. The presence of positron beams would then allow new tests that could confirm the constraints predicted by the EFT formalism or, alternatively, might indicate new light particles that would require further corrections outside this formalism. 

\section{Physics beyond the Standard Model}
\label{sec:bsm}
As opposed to the Higgs and SM precision program, which is guaranteed at the ILC, statements about new physics models could well remain speculative until we have ILC data in hand.  
While a discovery of course would be the ultimate triumph, also exclusion bounds can change the paradigms of our thinking. A recent example are the null-results in searches for plain vanilla SUSY as predicted by the constrained MSSM in the first run of the LHC. We have explained in~\cite{Fujii:2017vwa, Fujii:2017ekh} that the ILC, even at 250\,GeV, has an interesting window for new particle discovery consistent with the exclusions of new particles reported by the LHC experiments. However, still, these particles might not appear.

The role of positron polarization in BSM physics is sometimes discussed separately for discovery or exclusion, on one hand, and for characterization of the signal once it has been found, on the other hand, implying that e.g.\ in case of the ILC a polarized source could be built once a discovery has been made.

There are examples where this approach is valid, e.g.\ rather plain-vanilla SUSY, where e.g.\ pair production of (sufficiently light) scalar electrons could be disovered with a large significance after a few weeks or months of data taking~\cite{Berggren:2015qua}, even without positron polarization. Then, indeed a polarized source would be essential and could be added in order to perform a full study of the chiral structure of the new states~\cite{MoortgatPick:2005cw,Moortgat-Pick:redreport}.

In general, though, and especially for the highest accessible masses and the lowest detectable couplings, one could very well end up in a situation where a deviation from the SM becomes visible only with the full  dataset, possibly even with a medium significance between $3$ and $5\,\sigma$. In such cases, the ability to test the ``characterization'' part of the BSM program by exploiting the extra observables provided by positron polarization would give important and in some cases even crucial hints for convincing ourselves that the observed effect is indeed new physics, and would narrow down the possible interpretations. In other words, only in special cases is there a strict separation between ``discover first'' and ``characterize later'', while in real life, both are often intertwined. 


Thus, there are examples where a discovery of new particle might be missed without the assistance of positron polarization.  The most important of these are the following:

\begin{itemize}
\item {\bf Invisible particles:} A very important search in which $\ee$ colliders offer a distinct advantage is the search for pair production of invisible particles in association with initial-state radiation. This search can uncover WIMP dark matter particles and other very weakly coupled states. At an $\ee$ collider, this is the mono-photon (photon + missing energy) signature~\cite{Habermehl:2017dxh,Bartels:2012ex}. The SM predicts a low rate of mono-photon events, and that rate is calculable at the part per mille level.  The pair-production reaction has a definite chirality, so that it occurs either in the LR/RL or LL/RR chirality configurations.  Thus, positron polarization allows us to collect event samples in which the new physics effect is reduced and the SM backgrounds can be measured.

For the model of invisible particle production through an effective operator, the exclusion reach in the plane of the particle mass $M_\chi$ and the mediator scale $\Lambda$ has been evaluated for \Pp=30\% and \Pp=0 based on~\cite{Habermehl:2017dxh}. Since the cross section depends on the mediator scale as $\Lambda^{-4}$, the extension with positron polarization is significant, the equivalent of a 25\% increase in running time\footnote{With \Pp=0, an additional 940\,fb$^{-1}$ would be needed to reach the same sensitivity. Assuming the luminosity upgrade with an annual luminosity of 384\,fb$^{-1}$, this corresponds to 2.5 years of additional operation, which is about 25\% of the running time foreseen for the 250\,GeV stage.}. This does not account the ability to test the control of systematics, as described in the previous paragraph, or the fact that, in the event of a discovery, positron polarization can assist in determining the operator structure of the production process~\cite{Bartels:2012ex,Dreiner:2012xm,Chae:2012bq,Choi:2015zka}.

\item {\bf Heavy Leptons:} The study of $W$ pair production can reveal $t$-channel exchange of new heavy leptons with masses beyond the ILC collider energy. Exchange of such heavy leptons changes the chirality structure of the production process, so it is in principle distinguishable from modification of the triple gauge couplings, which leaves the production chirality-conserving. To measure this effect, however, double polarization asymmetries are needed~\cite{Moortgat-Pick:2013jra}.  Thus, for this search, positron polarization is required.



\item {\bf $R$-Parity Violating SUSY:} While there are strong bounds from the LHC on strongly interacting SUSY particles decaying into final states with large missing transverse energy (MET), electroweak new particles are much less constrained, especially in $R$-parity violating models, which give small MET. A search that is particularly powerful for $\ee$ colliders is that for $s$-channel exchange of a scalar neutrino, which would then decay to $\mu^+\mu^-$ or $\tau^+\tau^-$.  This is another example in which the new process has a chirality structure different from that of the SM. In this case, longitudinal positron polarization increases the signal-to-background ratio by more than a factor 2~\cite{Moortgat-Pick:redreport}. This leads to an increased discovery reach, not only to higher masses, but also to smaller R-parity violating couplings.



\item {\bf Contact Interactions:} Finally, a model-independent search for contact interactions in Bhabha scattering profits significantly from the ability to measure the cross sections for all four helicity combinations which are available with both beam polarized~\cite{Moortgat-Pick:redreport}. In this example, positron polarization increases the reach in probed energy scales of new physics by a factor~1.3.


\end{itemize}


\section{Conclusions}
\label{sec:conclusions}
In this document we discussed the impact of positron polarization on the ILC physics program, with special emphasis on a first energy stage at 250\,GeV. While the  {\em statistical} effects on many standard
Higgs and electroweak observables are small, and could be compensated by longer operation time, positron polarization plays an important role in controlling {\em systematic} uncertainties, and thus enables us to fully exploit the potential of the ILC. Without positron polarization, many important measurements will be limited by systematic uncertainties. 

In many cases positron polarization opens the door towards more model-independent interpretations of the data, which are important both in presence and in absence of discoveries. In case of discoveries, positron polarization plays a large role in identifying the underlying new physics model. But it also can be the decisive handle to identify an observation with (otherwise) medium significance as incompatible with the SM, and thus as a real discovery.

In conclusion, we find that positron polarization has an important role to play in the ILC program. 

\Acknowledgements
The authors thank the ILD and SiD detector concepts groups for providing material for this document. We thank Peter Rowson and Michael Woods for insight into the SLD experience discussed in Section~\ref{sec:polarimetry}.
T.~Barklow, M.~Berggren, C.~Grojean, M.~Habermehl, R.~Karl, J.~List, G.~Moortgat-Pick and J.~Reuter thankfully acknowledge the support by the Deutsche Forschungsgemeinschaft (DFG) through the Collaborative Research Centre SFB 676 Particles, Strings and the Early Universe, projects B1 and B11. C.~Grojean is also supported by the European Commission through the Marie Curie Career Integration Grant 631962 and by the Helmholtz Association. F.~Simon acknowledges the support of the DFG cluster of excellence `Origin and Structure of the Universe'. R.~P\"oschl is supported by the Quarks \& Leptons programme of the French IN2P3. G.~W.~Wilson is supported by the US National Science Foundation under Award PHY-1607262. The work of T.~Barklow and M.~Peskin is supported by the U.S. Department of Energy, contract
DE-AC02-76SF00515.
Some material is based upon work supported by the U.S. Department of Energy, Office of Science, 
Office of High Energy Physics under Award Number DE-SC-0009956.
This work is also supported by the Grants-in-Aid for Science Research No. 16H02173 and 16H02176 of the Japan Society for Promotion of Science (JSPS).

\end{document}